\documentclass[journal]{IEEEtran}
{\MakeLowercase{\textit{Song et al.}} {HemSeg-200}: A Voxel-Annotated Dataset for Intracerebral Hemorrhages Segmentation in Brain CT Scans.}
\usepackage{cite}
\usepackage{amsmath,amssymb,amsfonts}
\usepackage{algorithmic}
\usepackage{graphicx}
\usepackage{textcomp}
\usepackage{multirow}
\usepackage{url}
\usepackage{hyperref}
\usepackage{tabularx}
\usepackage{multirow}
\usepackage{array}
\usepackage{booktabs}
\usepackage{hhline}
\usepackage[top=0.75in, bottom=0.75in, left=0.75in, right=0.75in]{geometry}
\begin{document}
\title{{HemSeg-200}: A Voxel-Annotated Dataset for Intracerebral Hemorrhages Segmentation in Brain CT Scans}
% [Anonymous]
% \author{Anonymous authors} 
\author{Changwei Song, Qing Zhao, Jianqiang Li, Xin Yue, Ruoyun Gao, Zhaoxuan Wang, An Gao and Guanghui Fu\textsuperscript{*}
\thanks{Changwei Song, Jianqiang Li, Qing Zhao, Xin Yue, Ruoyun Gao, Zhaoxuan Wang are with School of Software Engineering, Beijing University of Technology, Beijing, China.}
\thanks{An Gao is with Department of Radiology, Tianjin Medical University Cancer Institute and Hospital, Tianjin, China.}
\thanks{Guanghui Fu is with Sorbonne Universit\'{e}, Institut du Cerveau – Paris Brain Institute - ICM, CNRS, Inria, Inserm, AP-HP, H\^{o}pital de la Piti\'{e}-Salp\^{e}tri\`{e}re, Paris, France (e-mail: guanghui.fu@inria.fr).}
\thanks{Corresponding author: Guanghui Fu (\url{guanghui.fu@inria.fr})}
\thanks{This work was supported by The National Natural Science Foundation of China under Grant No.72204169.}
\thanks{This work has been submitted to the IEEE for possible publication. Copyright may be transferred without notice, after which this version may no longer be accessible.}
}

%\newgeometry{top=1in, bottom=0.75in, left=0.75in, right=0.75in}
\maketitle
%\restoregeometry

\begin{abstract}
Acute intracerebral hemorrhage is a life-threatening condition that demands immediate medical intervention. Intraparenchymal hemorrhage (IPH) and intraventricular hemorrhage (IVH) are critical subtypes of this condition. Clinically, when such hemorrhages are suspected, immediate CT scanning is essential to assess the extent of the bleeding and to facilitate the formulation of a targeted treatment plan. While current research in deep learning has largely focused on qualitative analyses, such as identifying subtypes of cerebral hemorrhages, there remains a significant gap in quantitative analysis crucial for enhancing clinical treatments. Addressing this gap, our paper introduces a dataset comprising 222 CT annotations, sourced from the RSNA 2019 Brain CT Hemorrhage Challenge and meticulously annotated at the voxel level for precise IPH and IVH segmentation. This dataset was utilized to train and evaluate seven advanced medical image segmentation algorithms, with the goal of refining the accuracy of segmentation for these hemorrhages. Our findings demonstrate that this dataset not only furthers the development of sophisticated segmentation algorithms but also substantially aids scientific research and clinical practice by improving the diagnosis and management of these severe hemorrhages.
Our dataset and codes are available at \url{https://github.com/songchangwei/3DCT-SD-IVH-ICH}. 
\end{abstract}
\IEEEpeerreviewmaketitle

\section{Introduction}

\IEEEPARstart{I}{ntracerebral} hemorrhage (ICH) is one of the most lethal types of stroke, with a one-month mortality rate exceeding 40\%~\cite{brouwers2014predicting}. ICH encompasses various forms of bleeding within the skull, which can occur either outside or within the brain tissue~\cite{gebel2000intracerebral}. The quantification of intraparenchymal hemorrhage (IPH) and intraventricular hemorrhage (IVH) is crucial for clinical doctors, as these measurements are often particularly relevant to surgical planning~\cite{gross2019cerebral, hanley2009intraventricular}.
The high mortality rate associated with intracranial hemorrhage highlights the urgent need for research into innovative diagnostic methods. Diagnostic tools based on deep learning show great promise as critical tools to assist clinicians in accurately diagnosing such conditions.~\cite{yeo2021review,neethi2024comprehensive}.
For example, Fu et al.~\cite{fu2021attention} utilized convolutional neural networks with attention mechanisms to qualitatively classify intracranial hemorrhages. Their method was applied to five types of hemorrhages across the RSNA (RSNA Intracranial Hemorrhage Detection)~\cite{rsna_flanders2020construction, rsna_kaggle} and CQ500~\cite{cq500_chilamkurthy2018deep} datasets. They also provided interpretive analyses of their results by attention mechanism to highlight suspicious areas in the imaging data. 
Building on this work, Wang et al.~\cite{wang2022diagnosis} developed a method that replicates the reading behaviors of expert radiologists. Their approach involves selectively magnifying original images from the full sequence scans to focus on and diagnose suspicious lesions. 
They conducted human-computer interaction experiments which demonstrated that with the assistance of deep learning models, junior physicians improved their diagnostic capabilities. Specifically, these physicians increased their diagnostic accuracy by 16\% points F1-score, reduced their average diagnostic time by 10 seconds, and reported higher levels of confidence in their clinical assessments.
To date, most research in this area has focused on qualitative categorical studies, there has been limited focus on the problem of hemorrhage segmentation, with the majority of these studies utilizing private datasets~\cite{neethi2024comprehensive}. 
%Deep learning-based medical image segmentation significantly aids physicians in diagnosing diseases effectively, and the focus is increasingly shifting towards its application in intracerebral hemorrhage (ICH) segmentation. 
Deep learning-based medical image segmentation algorithms can automatically find anatomical structures or lesion areas, providing a foundation for quantitative disease research~\cite{ramesh2021review}.
Pioneering studies in this area have demonstrated a variety of innovative techniques. 
Yu et al.~\cite{yu2022robust} used dimensionality reduction in conjunction with the U-net model~\cite{ronneberger2015u} to segment and measure brain hemorrhages. Nijiati et al.~\cite{nijiati2022symmetric} improved the accuracy of segmentation and classification of ICH lesions by integrating the Transformer~\cite{vaswani2017attention} with the traditional U-Net~\cite{ronneberger2015u} using symmetry-based prior knowledge. In addition, Peng et al.~\cite{peng2022deep} advanced CT image segmentation and ICH volume measurement by incorporating attention gate mechanisms along with a focus structure for improved accuracy. Finally, Chang et al.~\cite{9948588} refined class-specific feature extraction using channel attention in the U-net encoder and employed both spatial and channel attention in the decoder to better delineate shapes and classify types.
Chang et al.~\cite{chang2022pesa} highlighted the importance of tissue deformation, such as perihematomal edema (PHE), as an indicator of hemorrhagic regions. They introduced a novel center-surround differential U-net (CSD U-net) that detects areas of tissue deformation as potential hemorrhagic sites while also predicting actual hemorrhage. Tong et al.~\cite{tong2023deep} improved the segmentation of intracerebral and intraventricular hemorrhage in CT images by integrating a 3D U-net with a multi-scale boundary awareness module, which improved the delineation of hematoma boundaries. Their model also includes a consistency loss feature to reduce the misclassification of pixels into multiple categories.
However, current research suggests that algorithmic performance improvements under identical data conditions are modest~\cite{isensee2024nnu}, with the emphasis increasingly shifting to dataset construction. Unfortunately, there are very few publicly available datasets for segmentation tasks, and most are neither comprehensive nor perfect~\cite{tajbakhsh2020embracing}.

For tasks related to identifying subtypes of brain hemorrhage, there are established datasets such as CQ500~\cite{cq500_chilamkurthy2018deep} and the RSNA 2019 Brain CT Hemorrhage Challenge dataset (referred to as the RSNA dataset)~\cite{rsna_flanders2020construction}. The CQ500 dataset includes 491 patients represented by 1,181 head CT scans, while the RSNA dataset includes a significantly larger cohort of 16,900 patients with 19,336 head CT scans, providing extensive resources for deep learning model development. However, neither dataset provides pixel/voxel-level annotations for hemorrhage region segmentation, which poses a challenge for detailed analysis and model training.
Currently, the PhysioNet~\cite{hssayeni2020computed} and INSTANCE22~\cite{li20222022} datasets are public resources for brain hemorrhage segmentation tasks. However, these datasets are limited in terms of sample size; the PhysioNet dataset contains 82 CT scans, while the INSTANCE22 dataset contains 130 CT scans. The limited availability of samples in public datasets for brain hemorrhage segmentation is primarily due to the labor-intensive and time-consuming process required for pixel-level annotation.

In this paper, we focus on the segmentation of intraparenchymal hemorrhage (IPH) and intraventricular hemorrhage (IVH) lesions that are useful for quantitative analysis by medical doctors. We annotated a dataset of 222 brain CT scans called HemSeg-200 (IPH: 114 scans, IVH: 108 scans). We selected the relevant data from the RSNA dataset and annotated them at the voxel level.
To address the labor-intensive and time-consuming nature of pixel-level annotation in 3D CT datasets, this study presents a semi-automated approach that progresses from coarse to fine-grained annotation. We have developed a multistep annotation process in which initial coarse segmentations are gradually refined using a combination of automated techniques and manual corrections, and experienced physicians corrected again the annotation to ensure the quality of the labels. We evaluated the performance of the dataset by evaluating seven SOTA 3D medical image segmentation models. The experimental results showed that the nnU-net model yielded the highest performance, achieving a Dice score of 78.99\%.
Our annotations and codes are openly accessible, providing valuable resources for the development of deep learning models within the community. 

\section{Datasets} \label{sec:datasets} 
\subsection{Semi-automated data annotation}  \label{sec:datasets:annotation} 
Our dataset originates from the RSNA dataset~\cite{rsna_flanders2020construction, rsna_kaggle}, which was initially stored in DICOM format. 
We utilized the ITK tools~\cite{mccormick2014itk} for the batch conversion of the DICOM data into the NIfTI format~\cite{larobina2014medical}.
We applied windowing techniques to enhance image contrast, ensuring the visibility and clarity of features relevant to our study. 
Furthermore, to augment the efficiency of our data annotation process, we adopted a semi-automated, coarse-to-fine method. 

The first round of annotations was generated by TotalSegmentator~\cite{wasserthal2023totalsegmentator}, a tool developed on top of the nnU-net framework~\cite{isensee2021nnu} and trained on a wide range of public and proprietary datasets. 
In the second round, we randomly selected 20 scans to be independently labeled by four junior annotators. The inter-rater variability ranged from 75.33\% to 81.24\% (mean 78.45\%) based on Dice scores. They continued their annotations under the supervision of an expert. Subsequently, all annotations were thoroughly reviewed and verified by a board-certified neuroradiologist (A.G.) with two years of experience in neuro-oncology imaging to ensure the accuracy and reliability of the labels.
Throughout this process, we utilized 3D Slicer~\cite{pieper20043d} as our annotation tool.

%\begin{table}[!ht]
%\centering
%\caption{The Dice scores between pairs of annotators are denoted by A, B, C, and D.}
%\label{tab:dice_comparison}
%\begin{tabular}{|c|c|}
%\hline
%Pairs of Annotators & Dice \\ \hline
%A-B  & 77.78\% \\ \hline
%A-C  & 81.25\% \\ \hline
%A-D  & 81.09\% \\ \hline
%B-C  & 75.33\% \\ \hline
%B-D  & 78.16\% \\ \hline
%C-D  & 77.06\% \\ \hline
%Average & 78.45\% \\ \hline
%\end{tabular}
%\end{table}

\subsection{Dataset information} \label{sec:datasets:information} 
\begin{figure*}[!hbtp]
\centering
\includegraphics[width=0.80\linewidth]{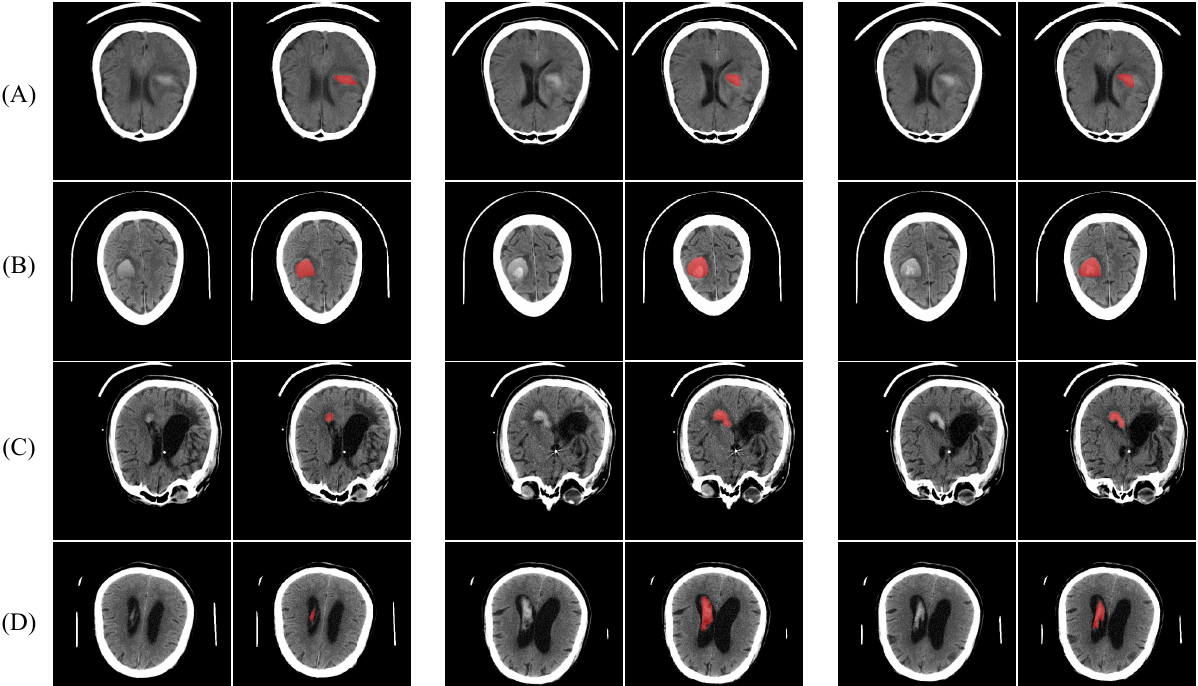}
\caption{Examples of intraparenchymal hemorrhage (IPH): case (A) and (B) and intraventricular hemorrhage (IVH): case (C) and (D). We shown the original CT slices and their overlaid annotations.}
\label{fig:ICH_case}
\end{figure*}

The dataset consists of 222 CT volumes, with 114 volumes diagnosed with intraparenchymal hemorrhage (IPH) and 108 volumes identified as intraventricular hemorrhage (IVH). 
All volumes share a consistent cross-sectional resolution of $512 \times 512$, derived from the standardized DICOM file format of the RSNA source. 
These CT scans contain different numbers of slices, from 24 to 56 (mean 33 slices). In-plane resolution ranges from 0.39 mm to 0.63 mm and slice thickness ranges from 3.00 mm to 6.50 mm.
Figure~\ref{fig:ICH_case} provides some examples of the dataset and overlap with its annotation.

\section{Methods} \label{sec:methods} 
We evaluated seven commonly used 3D medical image segmentation models in the field, which can facilitate the understanding of the performance of these commonly used algorithms on this dataset.
\begin{itemize}
    \item \textbf{3D U-net~\cite{cciccek20163d}:} 3D U-net features a symmetric U-shaped architecture, comprising a contracting (encoding) path and an expanding (decoding) path, designed to focus on preserving and recovering the fine structure of input images. The contracting path downsamples through standard convolution and pooling layers, while the expanding path upsamples through transpose convolutions. Skip connections directly concatenate feature maps from each layer of the contracting path with the corresponding layer in the decoding path, enhancing the preservation of details such as edges.
    \item \textbf{3D V-net~\cite{milletari2016v}:} V-net is a 3D fully convolutional neural network akin to U-net, utilizing an encoder-decoder structure with skip connections to merge low and high-level features. However, V-net differs by incorporating residual learning to facilitate gradient flow and enhance training stability; uses a combination of upsampling and convolution for feature map magnification, in contrast to U-net's typical use of transpose convolutions; and introduces a novel deep supervision mechanism that allows loss calculation at different network depths during training, aiding in more effective learning.
    \item \textbf{SegResNet~\cite{myronenko20193d}:} SegResNet consists of an encoder and a decoder, with the encoder employing $3\times3\times3$ convolutional kernels and GroupNorm normalization for downsampling. The decoder features two branches: one predicts the mask, and the other mirrors a variational autoencoder's decoder. Upsampling in the decoder is achieved through interpolation and $1\times1\times1$ convolutional kernels.
    \item \textbf{Attention U-net~\cite{oktay1804attention}:} Attention U-net enhances the standard U-net model with a novel attention gate module, integrated into the U-net's skip connections to heighten the model's sensitivity to foreground pixels.
    \item \textbf{UNETR~\cite{hatamizadeh2022unetr}:} UNETR merges the Transformer architecture with the U-Net design, featuring an encoder-decoder structure. In the encoder, it leverages self-attention to capture long-range dependencies, crucial for medical image segmentation where contextual information across the entire volume is vital for accurate delineation. The model also employs convolutional kernels to extract multi-scale features from various Transformer layers. In the decoder, it utilizes skip connections to merge these multi-scale features from the encoder with upsampled features in the decoder to enhance segmentation performance.
    \item \textbf{Swin UNETR~\cite{hatamizadeh2021swin}:} The Swin UNETR model extends the UNETR architecture by substituting its Transformer module with a Swin Transformer~\cite{liu2021swin}, employing a shifted window mechanism to compute self-attention across multiple resolutions for feature representation.
    \item \textbf{nnU-net~\cite{isensee2021nnu}:} nnU-Net is an automated framework built upon 2D U-Net, 3D U-Net, and cascaded U-Net models. It utilizes heuristic rules to tailor data preprocessing, augmentation, network architecture, and post-processing to the dataset, establishing a key benchmark in medical image segmentation.
\end{itemize}

We employed the Dice loss~\cite{milletari2016v,sudre2017generalised} as the loss function, which is commonly applied in medical image segmentation tasks. 

\section{Experiments and results} \label{sec:experiment} 

\subsection{Data preprocessing} \label{sec:experiment:preprocessing} 
% The preprocessing of medical images and their corresponding labels plays a pivotal role in ensuring consistency across the dataset and enhancing the performance of machine learning models. 
% This process involves several crucial steps to standardize the dataset. 
Initially, we aligned all images and masks with the RAS (Right, Anterior, Superior) coordinate system to establish a uniform orientation. 
Following this, we resampled both images and masks to unify the spatial resolution across each dimension according to the dataset's mean. 
For images, bilinear interpolation was utilized to maintain quality, whereas masks underwent nearest-neighbor interpolation to ensure the possibility of inverse transformations. 
Furthermore, we normalized the intensity values of the images. 
This entailed collecting intensity values for foreground classes (excluding background and ignored categories) from all training instances to compute the mean, standard deviation, and the 0.05 and 99.5 percentiles. 
The intensity values were then clipped to these percentile values, with subsequent subtraction of the mean and division by the standard deviation, applying this normalization method consistently to each training case for the respective input channel. 
Additionally, to augment data variability and boost the model's ability to generalize, we incorporated random flip transformations along the x and y axes. 
During the train and inference phase, we utilized a sliding window strategy, extracting patches of size $256 \times 256 \times 16$ for processing.
To ensure comprehensive coverage, four patches were extracted from each image, with at least one patch containing foreground elements.
Note that since we employed a sliding window strategy, there was no need to perform uniform operations such as resampling or resizing on CT images with inconsistent resolutions.
The dataset was stratified splited into training, validation, and testing subsets in a 3:1:1 ratio, resulting in 133, 44, and 45 volumes for each respective set. 

\subsection{Implementation details} \label{sec:experiment:implementation} 
Training and evaluation were conducted using a NVIDIA 24GB RTX 4090 GPU. 
Our model was developed using PyTorch framework~\cite{paszke2019pytorch}, and the MONAI library~\cite{cardoso2022monai}, tailored for medical image processing.
Regarding the model's hyperparameters, we opted for training over 200 epochs with a batch size of 2.
We employed the AdamW optimizer~\cite{kingma2014adam} with learning rate as 0.01 with a matching decay rate to ensure steady model convergence. 
In selecting performance metrics, we adhered to the guidelines set forth in Metrics Reloaded~\cite{maier2024metrics}, specifically employing the 3D Dice coefficient (Dice), intersection over union (IoU), and the 95\% 3D Hausdorff distance (HD95). For each metric, we report both the mean value and the corresponding 95\% confidence interval (CI), which was computed using bootstrap methods over the independent test set. 
We openly share our detailed configurations, source codes, and pre-trained model at: \url{https://github.com/songchangwei/3DCT-SD-IVH-ICH}.

\subsection{Results} \label{sec:results} 
\begin{figure*}[!htp]
\centering
\includegraphics[scale=0.85]{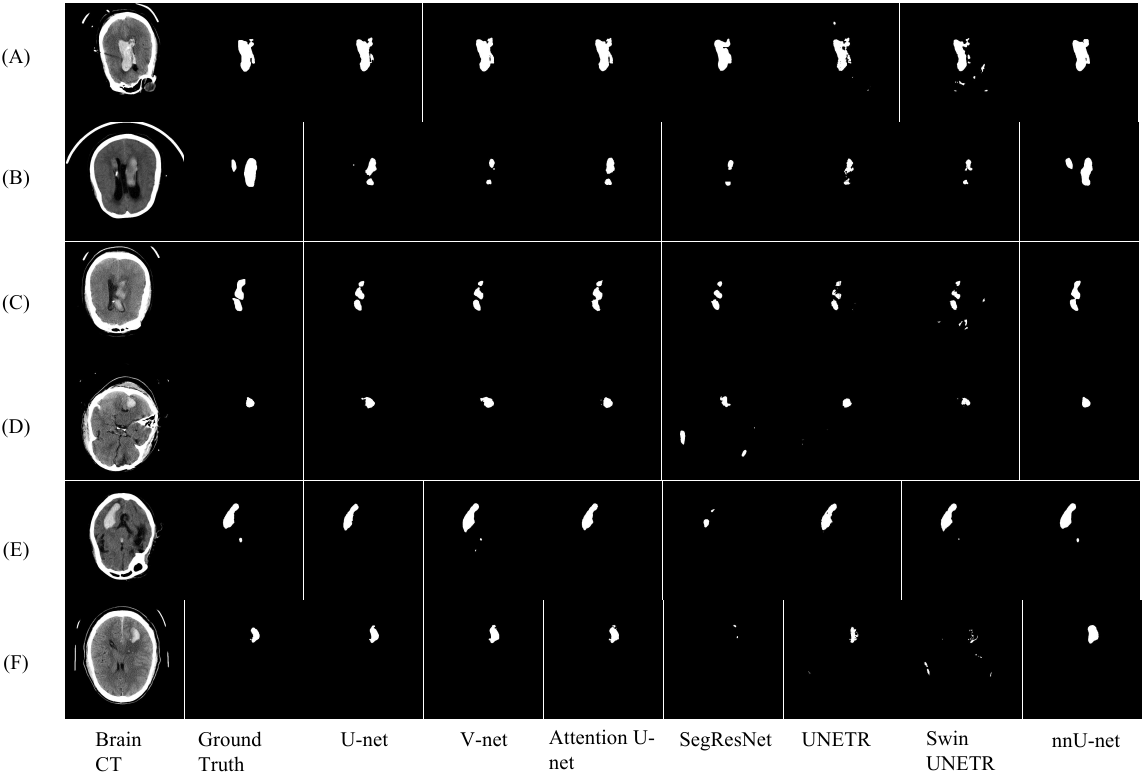}
\caption{Ground-truth label vs predictions on six representative patients with hemorrhages. }
\label{fig:result}
\end{figure*}

\begin{table*}[!ht]
\centering
\caption{The performance of IPH and IVH segmentation on brain CT scans from RSNA dataset. Results presented as mean with 95\% bootstrap confidence interval computed on the independent test set.}
\begin{tabular}{|l|c|c|c|}
\hline
\textbf{Models} & \textbf{Dice(\%)} & \textbf{IoU(\%)} & \textbf{HD95} \\ \hline
U-net           & 63.70 [55.82, 71.10]       & 51.40 [44.26, 58.48]    & 83.48 [51.66, 88.37]     \\ \hline
V-net           & 61.19 [52.81, 69.34]      & 49.39 [41.79, 56.98]     &   48.88 [33.30, 66.18]   \\ \hline
SegResNet       & 32.66 [25.96, 39.74]      & 22.02 [16.99, 27.57]    & 141.16 [124.20, 158.99]     \\ \hline
Attention U-net & 66.77 [58.90, 73.86]      & 54.98 [47.42, 62.10]    & 49.62 [33.44, 67.26]     \\ \hline
UNETR           & 55.98 [48.94, 62.69]      & 42.43 [36.10, 48.72]    & 111.10 [90.80, 131.54]     \\ \hline
Swin UNETR      & 48.50 [40.79, 56.13]      & 35.92 [29.19, 42.59]    & 164.85 [143.83, 186.99]     \\ \hline
nnU-net          &  \textbf{78.99 [73.21, 83.86]}      & \textbf{68.49 [62.05, 74.25]}      & \textbf{22.16 [9.19, 38.20]}     \\ \hline
\end{tabular}
\label{table:result}
\end{table*}
The experimental results, summarized in Table~\ref{table:result}, show that nnU-Net outperformed other models in segmentation performance on the dataset, achieving a Dice coefficient of 78.99\% with no overlap in the confidence interval with other models.
These superior results can be attributed to nnU-Net's comprehensive approach, which includes effective automated preprocessing, data augmentation, optimized network architecture, and postprocessing techniques.
In contrast, under identical data preprocessing conditions (as described in section~\ref{sec:experiment:preprocessing}), the Attention U-Net model performed best for Dice and IoU with values of 66.77\% and 54.98\%, respectively, while the V-Net model excelled in HD95 with a score of 48.88.
However, it was noted that Transformer-based models, specifically UNETR and Swin UNETR, did not perform as well on this dataset. 
% 添加Transformer需要更多数据来训练的参考文献
For a clearer comparison, Fig.~\ref{fig:result} displays the segmentation results from various models alongside ground truth.

The U-net architecture generally performs well, accurately identifying brain hemorrhage regions in most examples. However, finer details such as edges may not be well recognized, as seen in examples (A) and (C). In cases where multiple hemorrhages are present, U-net may miss some of the hemorrhages, as seen in examples (B), where the hemorrhage in the right ventricle is not detected, and (E) where the hemorrhage in the third ventricle.
For V-net, the performance follows a similar trend to U-net, although in example (B) most of the major hemorrhage is missing. This could be due to inconsistencies in the contrast of this bleed region, resulting in the inability of the model to accurately detect it.
Attention U-net achieves the second best performance next to nnU-net, but also misses the bleed in the right ventricle in example (B). SegResNet performs poorly, missing lesions as shown in examples (B), (E) and (F). In addition, it generates topology errors outside of the region as shown in example (D), suggesting that the model struggles to accurately detect the contrast of the hemorrhage and misidentifies high signal regions of the skull.
For the transformer-based architecture models, UNETR and Swin UNETR exhibit similar errors, often producing out-of-region errors as shown in examples (A), (C), and (F). This indicates their inability to capture and learn the features of the disease in this dataset, even though they cover most of the important regions.
Due to the principles and characteristics of brain hemorrhage, our data may exhibit a mixture of ICH and IVH, as shown in example~(C). Additionally, our dataset may include a mixture of new hemorrhage and old hemorrhage, as depicted in examples~(B) and~(C), which could represent stale brain hemorrhage. These characteristics pose challenges for the algorithm, leading to several models missing some regions.
However, nnU-net performs consistently well in these cases, with excessive segmentation observed only in some regions shown in example (F). This highlights the model's powerful ability to produce optimal results by learning from a combination of training strategies.

\section{Discussion} \label{sec:discussion}
Compared to the current brain hemorrhage segmentation dataset, our dataset has the most extensive annotations. However, the performance achieved by the models is still limited. While nnU-Net achieves relatively good performance, reaching a Dice score of 78.99\%, the  vision transformer-based framework fails to achieve satisfactory performance, suggesting the need for more data. Therefore, the next step is to continue the labeling process to develop larger datasets.

Although the best performance was achieved with nnU-net based on the auto-configuration method, we have not compared it with other models that use autoconfiguration, such as nnFormer~\cite{zhou2021nnformer} and CoTR~\cite{xie2021cotr}. Comparing it with a broader range of models is also one of our next steps.

\section{Conclusion} \label{sec:conclusion} 
Our study employed a semi-automatic annotation method to annotate 222 brain hemorrhage datasets obtained from the RSNA dataset. To our knowledge, this represents the largest publicly available dataset for brain hemorrhage segmentation studies. 
To evaluate the efficacy of deep learning algorithms on this dataset, we tested seven widely-used deep learning models. The findings indicate that nnU-net outperformed the others, achieving a Dice score of 78.99\% with confidence intervals that do not overlap with those of the other models.
This research effectively supports the advancement of deep learning models in this area, offering promising potential for future clinical applications.
\bibliographystyle{IEEEtran}
\bibliography{ref}

\begin{thebibliography}{10}
\providecommand{\url}[1]{#1}
\csname url@samestyle\endcsname
\providecommand{\newblock}{\relax}
\providecommand{\bibinfo}[2]{#2}
\providecommand{\BIBentrySTDinterwordspacing}{\spaceskip=0pt\relax}
\providecommand{\BIBentryALTinterwordstretchfactor}{4}
\providecommand{\BIBentryALTinterwordspacing}{\spaceskip=\fontdimen2\font plus
\BIBentryALTinterwordstretchfactor\fontdimen3\font minus \fontdimen4\font\relax}
\providecommand{\BIBforeignlanguage}[2]{{%
\expandafter\ifx\csname l@#1\endcsname\relax
\typeout{** WARNING: IEEEtran.bst: No hyphenation pattern has been}%
\typeout{** loaded for the language `#1'. Using the pattern for}%
\typeout{** the default language instead.}%
\else
\language=\csname l@#1\endcsname
\fi
#2}}
\providecommand{\BIBdecl}{\relax}
\BIBdecl

\bibitem{brouwers2014predicting}
H.~B. Brouwers, Y.~Chang, G.~J. Falcone, X.~Cai, A.~M. Ayres, T.~W. Battey, A.~Vashkevich, K.~A. McNamara, V.~Valant, K.~Schwab \emph{et~al.}, ``Predicting hematoma expansion after primary intracerebral hemorrhage,'' \emph{JAMA neurology}, vol.~71, no.~2, pp. 158--164, 2014.

\bibitem{gebel2000intracerebral}
J.~M. Gebel and J.~P. Broderick, ``Intracerebral hemorrhage,'' \emph{Neurologic clinics}, vol.~18, no.~2, pp. 419--438, 2000.

\bibitem{gross2019cerebral}
B.~A. Gross, B.~T. Jankowitz, and R.~M. Friedlander, ``Cerebral intraparenchymal hemorrhage: a review,'' \emph{Jama}, vol. 321, no.~13, pp. 1295--1303, 2019.

\bibitem{hanley2009intraventricular}
D.~F. Hanley, ``Intraventricular hemorrhage: severity factor and treatment target in spontaneous intracerebral hemorrhage,'' \emph{Stroke}, vol.~40, no.~4, pp. 1533--1538, 2009.

\bibitem{yeo2021review}
M.~Yeo, B.~Tahayori, H.~K. Kok, J.~Maingard, N.~Kutaiba, J.~Russell, V.~Thijs, A.~Jhamb, R.~V. Chandra, M.~Brooks \emph{et~al.}, ``Review of deep learning algorithms for the automatic detection of intracranial hemorrhages on computed tomography head imaging,'' \emph{Journal of neurointerventional surgery}, vol.~13, no.~4, pp. 369--378, 2021.

\bibitem{neethi2024comprehensive}
A.~Neethi, S.~K. Kannath, A.~A. Kumar, J.~Mathew, and J.~Rajan, ``A comprehensive review and experimental comparison of deep learning methods for automated hemorrhage detection,'' \emph{Engineering Applications of Artificial Intelligence}, vol. 133, p. 108192, 2024.

\bibitem{fu2021attention}
G.~Fu, J.~Li, R.~Wang, Y.~Ma, and Y.~Chen, ``Attention-based full slice brain ct image diagnosis with explanations,'' \emph{Neurocomputing}, vol. 452, pp. 263--274, 2021.

\bibitem{rsna_flanders2020construction}
A.~E. Flanders, L.~M. Prevedello, G.~Shih, S.~S. Halabi, J.~Kalpathy-Cramer, R.~Ball, J.~T. Mongan, A.~Stein, F.~C. Kitamura, M.~P. Lungren \emph{et~al.}, ``Construction of a machine learning dataset through collaboration: the rsna 2019 brain ct hemorrhage challenge,'' \emph{Radiology: Artificial Intelligence}, vol.~2, no.~3, p. e190211, 2020.

\bibitem{rsna_kaggle}
\BIBentryALTinterwordspacing
S.~M. Anouk, W.~Carol, C.~Chris, S.~George, K.-C. Jayashree, E.~k. Julia, P.~Luciano, K.~M. Marc, L.~Matt, C.~Phil, B.~Robyn, and H.~M. Safwan, ``Rsna intracranial hemorrhage detection,'' 2019. [Online]. Available: \url{https://kaggle.com/competitions/rsna-intracranial-hemorrhage-detection}
\BIBentrySTDinterwordspacing

\bibitem{cq500_chilamkurthy2018deep}
S.~Chilamkurthy, R.~Ghosh, S.~Tanamala, M.~Biviji, N.~G. Campeau, V.~K. Venugopal, V.~Mahajan, P.~Rao, and P.~Warier, ``Deep learning algorithms for detection of critical findings in head ct scans: a retrospective study,'' \emph{The Lancet}, vol. 392, no. 10162, pp. 2388--2396, 2018.

\bibitem{wang2022diagnosis}
R.~Wang, G.~Fu, J.~Li, and Y.~Pei, ``Diagnosis after zooming in: A multilabel classification model by imitating doctor reading habits to diagnose brain diseases,'' \emph{Medical physics}, vol.~49, no.~11, pp. 7054--7070, 2022.

\bibitem{ramesh2021review}
K.~Ramesh, G.~K. Kumar, K.~Swapna, D.~Datta, and S.~S. Rajest, ``A review of medical image segmentation algorithms,'' \emph{EAI Endorsed Transactions on Pervasive Health and Technology}, vol.~7, no.~27, pp. e6--e6, 2021.

\bibitem{yu2022robust}
N.~Yu, H.~Yu, H.~Li, N.~Ma, C.~Hu, and J.~Wang, ``A robust deep learning segmentation method for hematoma volumetric detection in intracerebral hemorrhage,'' \emph{Stroke}, vol.~53, no.~1, pp. 167--176, 2022.

\bibitem{ronneberger2015u}
O.~Ronneberger, P.~Fischer, and T.~Brox, ``U-net: Convolutional networks for biomedical image segmentation,'' in \emph{Medical image computing and computer-assisted intervention--MICCAI 2015: 18th international conference, Munich, Germany, October 5-9, 2015, proceedings, part III 18}.\hskip 1em plus 0.5em minus 0.4em\relax Springer, 2015, pp. 234--241.

\bibitem{nijiati2022symmetric}
M.~Nijiati, A.~Tuersun, Y.~Zhang, Q.~Yuan, P.~Gong, A.~Abulizi, A.~Tuoheti, A.~Abulaiti, and X.~Zou, ``A symmetric prior knowledge based deep learning model for intracerebral hemorrhage lesion segmentation,'' \emph{Frontiers in Physiology}, vol.~13, p. 977427, 2022.

\bibitem{vaswani2017attention}
A.~Vaswani, N.~Shazeer, N.~Parmar, J.~Uszkoreit, L.~Jones, A.~N. Gomez, {\L}.~Kaiser, and I.~Polosukhin, ``Attention is all you need,'' \emph{Advances in neural information processing systems}, vol.~30, 2017.

\bibitem{peng2022deep}
Q.~Peng, X.~Chen, C.~Zhang, W.~Li, J.~Liu, T.~Shi, Y.~Wu, H.~Feng, Y.~Nian, and R.~Hu, ``Deep learning-based computed tomography image segmentation and volume measurement of intracerebral hemorrhage,'' \emph{Frontiers in Neuroscience}, vol.~16, p. 965680, 2022.

\bibitem{9948588}
C.~Shuo~Chang, T.~Sheuan~Chang, J.~Lin~Yan, and L.~Ko, ``All attention u-net for semantic segmentation of intracranial hemorrhages in head ct images,'' in \emph{2022 IEEE Biomedical Circuits and Systems Conference (BioCAS)}, 2022, pp. 600--604.

\bibitem{chang2022pesa}
J.~Chang, I.~Choi, and M.~Lee, ``Pesa r-cnn: Perihematomal edema guided scale adaptive r-cnn for hemorrhage segmentation,'' \emph{IEEE Journal of Biomedical and Health Informatics}, vol.~27, no.~1, pp. 397--408, 2022.

\bibitem{tong2023deep}
G.~Tong, X.~Wang, H.~Jiang, A.~Wu, W.~Cheng, X.~Cui, L.~Bao, R.~Cai, and W.~Cai, ``A deep learning model for automatic segmentation of intraparenchymal and intraventricular hemorrhage for catheter puncture path planning,'' \emph{IEEE journal of biomedical and health informatics}, 2023.

\bibitem{isensee2024nnu}
F.~Isensee, T.~Wald, C.~Ulrich, M.~Baumgartner, S.~Roy, K.~Maier-Hein, and P.~F. Jaeger, ``nnu-net revisited: A call for rigorous validation in 3d medical image segmentation,'' \emph{arXiv preprint arXiv:2404.09556}, 2024.

\bibitem{tajbakhsh2020embracing}
N.~Tajbakhsh, L.~Jeyaseelan, Q.~Li, J.~N. Chiang, Z.~Wu, and X.~Ding, ``Embracing imperfect datasets: A review of deep learning solutions for medical image segmentation,'' \emph{Medical image analysis}, vol.~63, p. 101693, 2020.

\bibitem{hssayeni2020computed}
M.~Hssayeni, M.~Croock, A.~Salman, H.~Al-khafaji, Z.~Yahya, and B.~Ghoraani, ``Computed tomography images for intracranial hemorrhage detection and segmentation,'' \emph{Intracranial hemorrhage segmentation using a deep convolutional model. Data}, vol.~5, no.~1, p.~14, 2020.

\bibitem{li20222022}
X.~Li, K.~Wang, J.~Liu, H.~Wang, M.~Xu, and X.~Liang, ``The 2022 intracranial hemorrhage segmentation challenge on non-contrast head ct (ncct),'' 2022.

\bibitem{mccormick2014itk}
M.~McCormick, X.~Liu, J.~Jomier, C.~Marion, and L.~Ibanez, ``Itk: enabling reproducible research and open science,'' \emph{Frontiers in neuroinformatics}, vol.~8, p.~13, 2014.

\bibitem{larobina2014medical}
M.~Larobina and L.~Murino, ``Medical image file formats,'' \emph{Journal of digital imaging}, vol.~27, pp. 200--206, 2014.

\bibitem{wasserthal2023totalsegmentator}
J.~Wasserthal, H.-C. Breit, M.~T. Meyer, M.~Pradella, D.~Hinck, A.~W. Sauter, T.~Heye, D.~T. Boll, J.~Cyriac, S.~Yang \emph{et~al.}, ``Totalsegmentator: Robust segmentation of 104 anatomic structures in ct images,'' \emph{Radiology: Artificial Intelligence}, vol.~5, no.~5, 2023.

\bibitem{isensee2021nnu}
F.~Isensee, P.~F. Jaeger, S.~A. Kohl, J.~Petersen, and K.~H. Maier-Hein, ``nnu-net: a self-configuring method for deep learning-based biomedical image segmentation,'' \emph{Nature methods}, vol.~18, no.~2, pp. 203--211, 2021.

\bibitem{pieper20043d}
S.~Pieper, M.~Halle, and R.~Kikinis, ``3d slicer,'' in \emph{2004 2nd IEEE international symposium on biomedical imaging: nano to macro (IEEE Cat No. 04EX821)}.\hskip 1em plus 0.5em minus 0.4em\relax IEEE, 2004, pp. 632--635.

\bibitem{cciccek20163d}
{\"O}.~{\c{C}}i{\c{c}}ek, A.~Abdulkadir, S.~S. Lienkamp, T.~Brox, and O.~Ronneberger, ``3d u-net: learning dense volumetric segmentation from sparse annotation,'' in \emph{Medical Image Computing and Computer-Assisted Intervention--MICCAI 2016: 19th International Conference, Athens, Greece, October 17-21, 2016, Proceedings, Part II 19}.\hskip 1em plus 0.5em minus 0.4em\relax Springer, 2016, pp. 424--432.

\bibitem{milletari2016v}
F.~Milletari, N.~Navab, and S.-A. Ahmadi, ``V-net: Fully convolutional neural networks for volumetric medical image segmentation,'' in \emph{2016 fourth international conference on 3D vision (3DV)}.\hskip 1em plus 0.5em minus 0.4em\relax Ieee, 2016, pp. 565--571.

\bibitem{myronenko20193d}
A.~Myronenko, ``3d mri brain tumor segmentation using autoencoder regularization,'' in \emph{Brainlesion: Glioma, Multiple Sclerosis, Stroke and Traumatic Brain Injuries: 4th International Workshop, BrainLes 2018, Held in Conjunction with MICCAI 2018, Granada, Spain, September 16, 2018, Revised Selected Papers, Part II 4}.\hskip 1em plus 0.5em minus 0.4em\relax Springer, 2019, pp. 311--320.

\bibitem{oktay1804attention}
O.~Oktay, J.~Schlemper, L.~L. Folgoc, M.~Lee, M.~Heinrich, K.~Misawa, K.~Mori, S.~McDonagh, N.~Y. Hammerla, B.~Kainz \emph{et~al.}, ``Attention u-net: Learning where to look for the pancreas. arxiv 2018,'' \emph{arXiv preprint arXiv:1804.03999}, 1804.

\bibitem{hatamizadeh2022unetr}
A.~Hatamizadeh, Y.~Tang, V.~Nath, D.~Yang, A.~Myronenko, B.~Landman, H.~R. Roth, and D.~Xu, ``Unetr: Transformers for 3d medical image segmentation,'' in \emph{Proceedings of the IEEE/CVF winter conference on applications of computer vision}, 2022, pp. 574--584.

\bibitem{hatamizadeh2021swin}
A.~Hatamizadeh, V.~Nath, Y.~Tang, D.~Yang, H.~R. Roth, and D.~Xu, ``Swin unetr: Swin transformers for semantic segmentation of brain tumors in mri images,'' in \emph{International MICCAI Brainlesion Workshop}.\hskip 1em plus 0.5em minus 0.4em\relax Springer, 2021, pp. 272--284.

\bibitem{liu2021swin}
Z.~Liu, Y.~Lin, Y.~Cao, H.~Hu, Y.~Wei, Z.~Zhang, S.~Lin, and B.~Guo, ``Swin transformer: Hierarchical vision transformer using shifted windows,'' in \emph{Proceedings of the IEEE/CVF international conference on computer vision}, 2021, pp. 10\,012--10\,022.

\bibitem{sudre2017generalised}
C.~H. Sudre, W.~Li, T.~Vercauteren, S.~Ourselin, and M.~Jorge~Cardoso, ``Generalised dice overlap as a deep learning loss function for highly unbalanced segmentations,'' in \emph{Deep Learning in Medical Image Analysis and Multimodal Learning for Clinical Decision Support: Third International Workshop, DLMIA 2017, and 7th International Workshop, ML-CDS 2017, Held in Conjunction with MICCAI 2017, Qu{\'e}bec City, QC, Canada, September 14, Proceedings 3}.\hskip 1em plus 0.5em minus 0.4em\relax Springer, 2017, pp. 240--248.

\bibitem{paszke2019pytorch}
A.~Paszke, S.~Gross, F.~Massa, A.~Lerer, J.~Bradbury, G.~Chanan, T.~Killeen, Z.~Lin, N.~Gimelshein, L.~Antiga \emph{et~al.}, ``Pytorch: An imperative style, high-performance deep learning library,'' \emph{Advances in neural information processing systems}, vol.~32, 2019.

\bibitem{cardoso2022monai}
M.~J. Cardoso, W.~Li, R.~Brown, N.~Ma, E.~Kerfoot, Y.~Wang, B.~Murrey, A.~Myronenko, C.~Zhao, D.~Yang \emph{et~al.}, ``Monai: An open-source framework for deep learning in healthcare,'' \emph{arXiv preprint arXiv:2211.02701}, 2022.

\bibitem{kingma2014adam}
D.~P. Kingma and J.~Ba, ``Adam: A method for stochastic optimization,'' \emph{arXiv preprint arXiv:1412.6980}, 2014.

\bibitem{maier2024metrics}
L.~Maier-Hein, A.~Reinke, P.~Godau, M.~D. Tizabi, F.~Buettner, E.~Christodoulou, B.~Glocker, F.~Isensee, J.~Kleesiek, M.~Kozubek \emph{et~al.}, ``Metrics reloaded: recommendations for image analysis validation,'' \emph{Nature methods}, pp. 1--18, 2024.

\bibitem{zhou2021nnformer}
H.-Y. Zhou, J.~Guo, Y.~Zhang, L.~Yu, L.~Wang, and Y.~Yu, ``nnformer: Interleaved transformer for volumetric segmentation,'' \emph{arXiv preprint arXiv:2109.03201}, 2021.

\bibitem{xie2021cotr}
Y.~Xie, J.~Zhang, C.~Shen, and Y.~Xia, ``Cotr: Efficiently bridging cnn and transformer for 3d medical image segmentation,'' in \emph{Medical Image Computing and Computer Assisted Intervention--MICCAI 2021: 24th International Conference, Strasbourg, France, September 27--October 1, 2021, Proceedings, Part III 24}.\hskip 1em plus 0.5em minus 0.4em\relax Springer, 2021, pp. 171--180.

\end{thebibliography}
\end{document}